\documentclass[preprint]{iucr}              
\usepackage{url}

     \journalcode{J}              

\begin{document}                  



\title{McSAS3: improved Monte Carlo small-angle scattering analysis software for dilute and dense scatterers}


\cauthor[a]{Brian Richard}{Pauw}{brian.pauw@bam.de} 
 
\author[a]{Ingo}{Bre{\ss}ler} 


\aff[a]{Bundesanstalt f\"ur Materialforschung und -Pr\"ufung, Unter den Eichen 87, 12205 Berlin \country{Germany}}


\shortauthor{Pauw and Bre{\ss}ler}






\maketitle                        

\begin{synopsis}

McSAS3 consists of a pair of tools: a command-line, Monte Carlo-based fitting engine for headless, automated analysis of large quantities of small-angle scattering data, and a graphical user interface (GUI) to simplify configuration and batch fitting. It is the completely refactored successor to the McSAS software. 

\end{synopsis}

\begin{abstract}

McSAS3 is the refactored successor to the original McSAS Monte Carlo small-angle scattering analysis software. It is intended to be integrated in automated data processing pipelines, but can also be used to process individual (batches of) scattering data. 

McSAS3 comes with a graphical user interface (McSAS3GUI), complete with guides, examples and videos. McSAS3GUI will help to generate and test the three configuration files that McSAS3 needs for data read-in, Monte Carlo optimization and histogramming. The user interface can also be used to process individual files or batches, and can be augmented with machine-specific use templates.

The Monte Carlo (MC) approach is able to fit most practical scattering patterns extremely well, resulting in form-free model parameter distributions. Theoretically, these can be distributions on any model parameter, but in practice the MC-optimized parameter is usually a (volume-weighted) size distribution, in absolute volume fraction for absolute-scaled data. 

\end{abstract}

\section{Introduction}

Assuming a proper automated data correction pipeline is in place (c.f. \cite{Pauw-2017}), data analysis is often the most time-consuming step in a small-angle scattering experiment. With classical least-squares analysis (e.g. using SasView \cite{SasView-2025} or SASfit \cite{Kohlbrecher-2022}), time is needed to create a sample-appropriate model, consisting of a scatterer model, parameter distribution form, and inter-structure scattering descriptions. This triad then needs adapting and augmentation so it is able to match the peculiarities of real-life morphologies and measurements (a good example of which is shown in \cite{Henning-2022}). This becomes much more complex in dynamic experiments where the structure may evolve beyond the model during the experiment. 

The Monte Carlo method offers a significant advantage here, in that the parameter distribution \textit{form} no longer needs to be specified \cite{Pauw-2013a}. Only the overall scatterer morphology (``form factor''), and optionally the inter-particle scattering (``structure factor'') now need to be defined. As fewer choices means less chance for human bias in the result \cite{Pauw-2023}, reproducibility is improved. Once the form factor and optional structure factor are configured in accordance with the overall sample morphology, the optimization can start. 

The MC procedure will then attempt to match the measured data ($I_meas$) with the MC data ($I_mc$). This $I_mc$ is comprised of an inverse volume-weighted sum of independent, individual model contributions $I_\mathrm{contrib}$. The optimization is performed by randomly varying a specific parameter of the $n_\mathrm{contrib}\approx 300$ individual model contributions within predefined limits, and selecting the new parameter value when the match improves. The resulting parameter distributions are often less "smooth" as with the parametrized distribution shapes of the classical models (such as normal, log-normal or Schultz distributions), but can more closely describe the reality of the structure in a greater range of samples. 

Over the last decade, the original McSAS \cite{Bressler-2015} has seen widespread use, with its accompanying paper having been cited over 200 times already. One of the first examples demonstrating its utility was to describe the complex size distribution evolution during the synthesis of quantum dots, which allowed the reaction mechanism to be identified \cite{Abecassis-2015}. Other applications of the software include studying defect formation in titanium alloys \cite{Wu-2025}, condensed phases in magnetic fluids \cite{Mamiya-2021}, nanoparticles synthesised by laser ablation \cite{Letzel-2017}, liquid metal droplets in composites \cite{Crater-2022} and pores in energy storage materials \cite{Qiu-2025}. The vast majority use McSAS as a convenient tool for minimal-assumption nanostructural sizing, usually in combination with other methods such as electron microscopy for comparability and scatterer shape verification. 

\begin{figure}
\caption{The panels of the McSAS3GUI user interface showing an optimization on a ZIF-8 dataset, with the main UI shown in the top left, the loaded, clipped, rebinned data and fit shown in top right, the output PDF in the bottom left with the data and size distribution, and the evolution of the optimization parameter in the bottom right.}
\includegraphics[width=0.99\textwidth,trim={.5mm 3mm 1mm .5mm},clip]{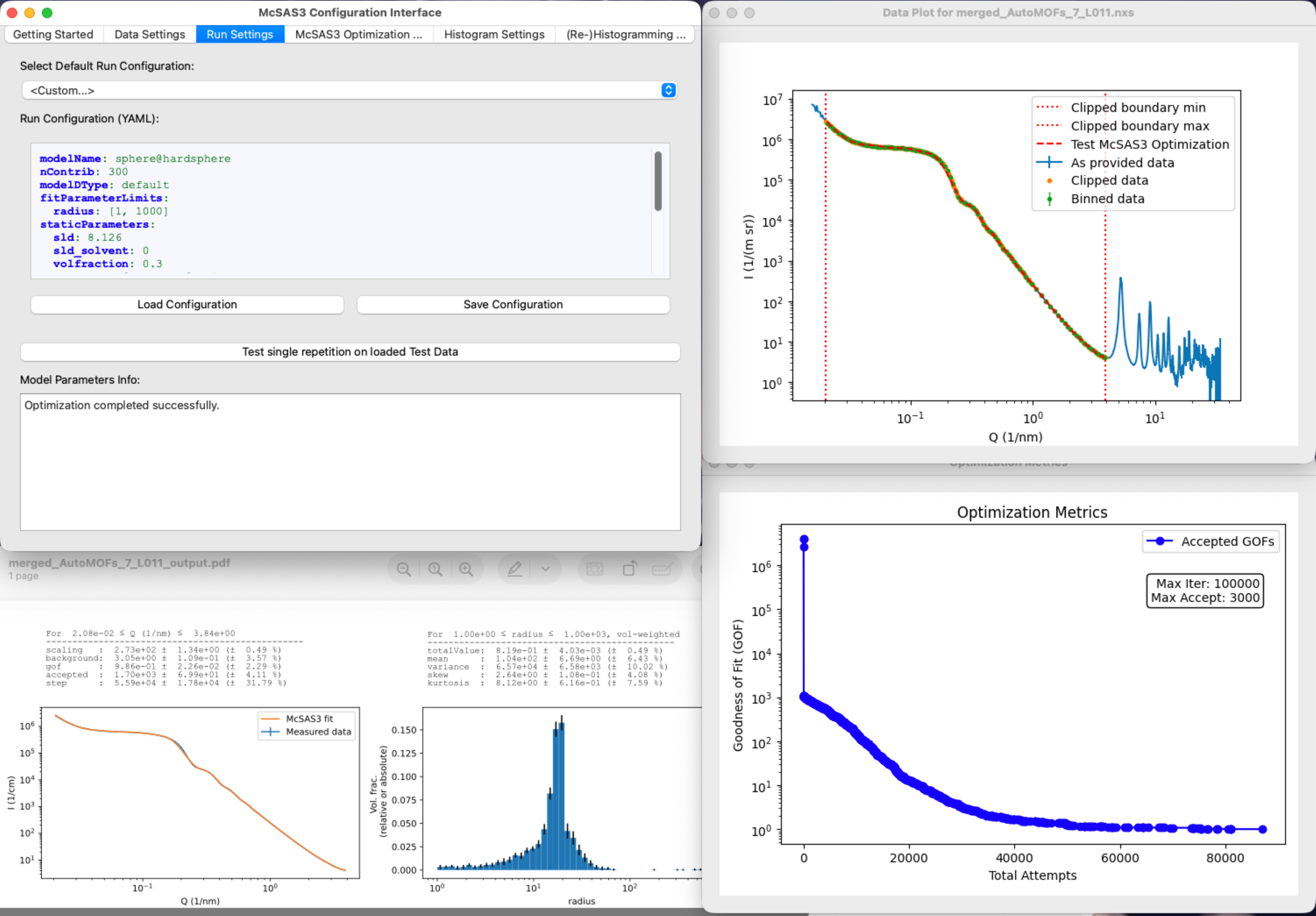}
\label{fg:UI}
\end{figure}

However, the original McSAS had issues that were impossible to resolve without a major rewrite. For example, the UI was inseparable from the core, preventing headless operation or integration in automated data processing pipelines. Secondly, the number of supported models was rather limited, and it did not support multicore processing. Most frustratingly from a usage perspective, you could not change the histogram settings without having to redo the complete (and occasionally time-consuming) optimization. Therefore, McSAS3\protect\footnote{after a brief but failed attempt at McSAS2...} was written as a complete rewrite of the original code. While software is never truly finished, McSAS3 + GUI, shown in Figure \ref{fg:UI}, now exceeds the ``good enough'' state where it is usable by the wider public. So far, McSAS3 has been used (amongst others) on large quantities of data from operando electrochemistry experiments \cite{Jansske-2025}, and a wide range of data from different distributions of metal-organic framework samples \cite{Smales-2025}, all using single, consistent models per application. 

Here, we will consider the original McSAS papers as read (in particular the underlying optimization principles \cite{Pauw-2013a, Bressler-2015}), and only highlight the improvements that McSAS3 + GUI bring. Some of the improvements will be demonstrated by three practical examples. 

\section{Advantages of McSAS3}

The Python-based McSAS3 (command-line) core offers the following advantages: 
\begin{itemize}
\item The McSAS3 core is designed to be integrated in automated data workflows, and can be executed from the command-line or a Jupyter notebook using configuration files. This scriptability allows for immediate, in-line analysis of data during an experiment as well. 
\item Input data is supported from either ASCII or HDF5 containers. Data read-in options can be specified in detail using a YAML configuration file, to adapt to a wide range of variants such as CSV, NeXus, NeXish \protect{\footnote{Looks like NeXus but isn't quite conforming to the standard. This is implemented at most synchrotron beamlines as well as our MOUSE instrument.}} and Anton Paar's (now superseded) ``PDH'' format. 
\item The main models library is SasModels \cite{SasModels-2025} from the SasView\cite{SasView-2025} community. This means that many (compiled) models are supported, and can be used in combination with some structure factors. The only limitation is that the chosen form factor model \emph{must specify a volume}. Combinations of models can even be constructed using a simple syntax (e.g. "sphere@hardspere+cylinder"), although it is recommended to keep models as simple as possible to avoid ambiguous results. 
\item There is a possibility to use simulated data as a model, as demonstrated in Paragraph \ref{ssc:ex3}. This can be helpful for scatterer morphologies that are unsupported by SasModels. 
\item Multi-threaded processing is implemented for the independent repetitions, making sure you can now use modern computers as a space heater again. Each process writes its intermediate results directly to the output HDF5 container.
\item You can now re-histogram any completed optimization to your heart's desire, without having to redo the optimization. 
\item McSAS3 was built in a very modular fashion, making it more maintainable, upgradeable and extensible.
\end{itemize}

While this is a good core which has been used already for several projects, we have recently developed a separate PyQT6-based user interface (McSAS3GUI) that can help adoption by lowering the barrier of use of McSAS3. The user interface helps with the generation of the three configuration files needed for the core (one for data read-in, one for the model and optimization settings, and one for the histogramming settings), and has tabs to execute batch fitting and batch histogramming. Furthermore, it has a user guide with templated examples, that can be extended to include user-specific templates for projects or bespoke configurations. Video guides are also available that explain the underlying principles, and demonstrate the use of McSAS3 with some general tips and tricks.

\section{Adminstrative details}

The code and issue lists are available on Github for the core\cite{mcsas3github} and for the user interface separately\cite{mcsas3guigithub}.
By using ``pip install McSAS3GUI'' in an appropriate Python environment, the core McSAS3 is installed automatically as a dependency. 
The code is in active development and will be supported for at least several years, resources permitting. The core and UI are released under an MIT license. A CI/CD pipeline has been implemented with a test suite, and contributions are very welcome using pull requests.

\section{Use examples}
\subsection{Use example 1: Data Analysis Round Robin bimodal gold nanoparticle suspension}

The first example shows how McSAS3 can be used to determine the modality of distributions, using data from the small-angle scattering data analysis round robin experiment \cite{Pauw-2023}. This was an experiment to find out how the results of a small-angle scattering analysis may depend on the person who is carrying out the analysis. To this end, we collated four datasets of dispersions of spherical scatterers, and provided: 
\begin{enumerate}
    \item the four three-column text files with the data, 
    \item A document with information on the units, the sample composition and photon energy used for the collection of the data 
    \item An excel sheet in which the researcher could fill in their results
\end{enumerate}
The results of this were published in \citeasnoun{Pauw-2023}, and the original information and datasets sent out to participants can be found in \citeasnoun{DARR-2023}. 
This example uses the first dataset, which is a measurement of a bimodal gold nanoparticle suspension. The smaller population was half the size and only 10\% in quantity of the larger population, to ensure the signal of the smaller particles are sufficiently hidden. 

This dataset is included as an example in McSAS3, which can fit this dataset with minimal effort\protect\footnote{After starting McSAS3GUI, select the ``round\_robin\_dataset\_1.yaml'' example/template, go to the ``McSAS3 Optimization...'' tab, click the ``Run McSAS3 Optimization...'' button, then go to the ``Histogram Settings'' tab, and click ``Test Histogramming''. As this is just one file, this test will generate the one PDF file and open it. The ``(Re-)Histogramming...'' tab will do the same, but not open the PDF file, and is more intended for histogramming batches of files.}. The ten independent optimizations of this example are completed in 15 seconds on a MacBook Pro M1, with the histogramming taking about a second more. 

\begin{figure}
\begin{center}
\includegraphics[width=0.99\textwidth,trim={0 0 30.2cm 0},clip]{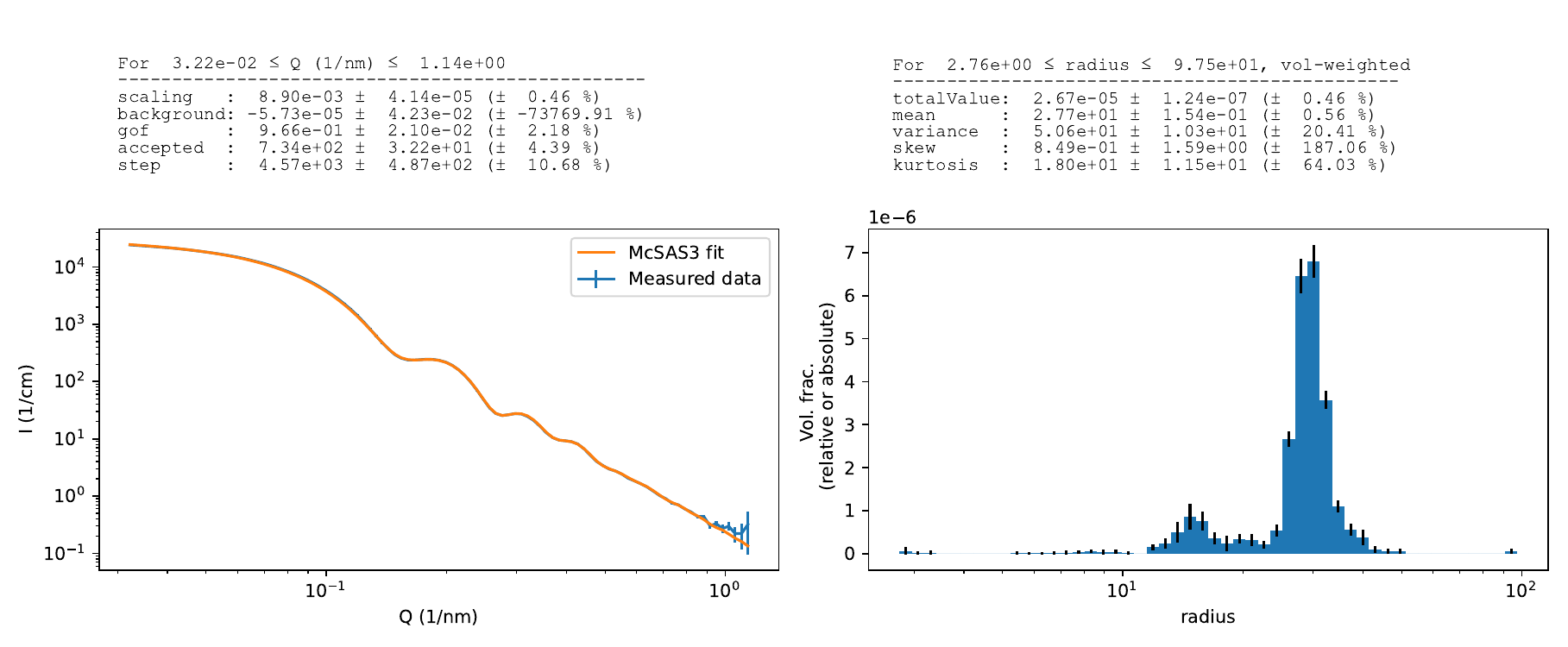}
\includegraphics[width=0.99\textwidth,trim={30.4cm 0 0 0},clip]{images/round_robin_dataset_1_output.pdf}
\caption{The analysis result of scattering from a bimodal gold nanoparticle suspension. Top left: data and fit, Top right: overall size histogram. Bottom left: first population, Bottom right: second population. Note that the lower graphs use linear binning edges, and therefore show a slightly different histogram than when using logarithmically-spaced histogram bin edges (as used in the top right figure).}
\label{fg:ex1}
\end{center}
\end{figure}

The resulting histogram (Figure \ref{fg:ex1}) shows the data and the fit in the leftmost panel, then the overall radius distribution, then just the small population, and lastly the large population. Population statistics showing the total volume fraction and volume-weighted population parameters (mean, variance, skew and kurtosis) of each individual range is shown above the individual histograms. It is recommended to express the population distribution width as a standard deviation by taking the square root of the variance. 

\subsection{Use example 2: Data Analysis Round Robin bimodal silica powder}

The second example demonstrates the use of McSAS3 on powder samples. It uses the third dataset of the aforementioned Data Analysis Round Robin, which contains a measurement of a powder mixture of two vastly different sizes of spherical silica particles. As this is an intercalating powder of low-polydispersity particles, there is a significant structure factor present. 

For data like this example, if we have significant structure factors and large structure scattering, we want to increase the maximum dimension in the model size parameter to ensure we can easily get an $I \propto q^{-4}$ Porod-slope from the smallest Q. Secondly, We also need to limit the data range in the data read-in configuration to avoid the wide-angle silica scattering hump. Furthermore, we increased the number of contributions to reduce the granularity of the resulting parameter distributions at the cost of increased computation time. 

\begin{figure}
\begin{center}
\includegraphics[width=0.99\textwidth,trim={0 0 31cm 0},clip]{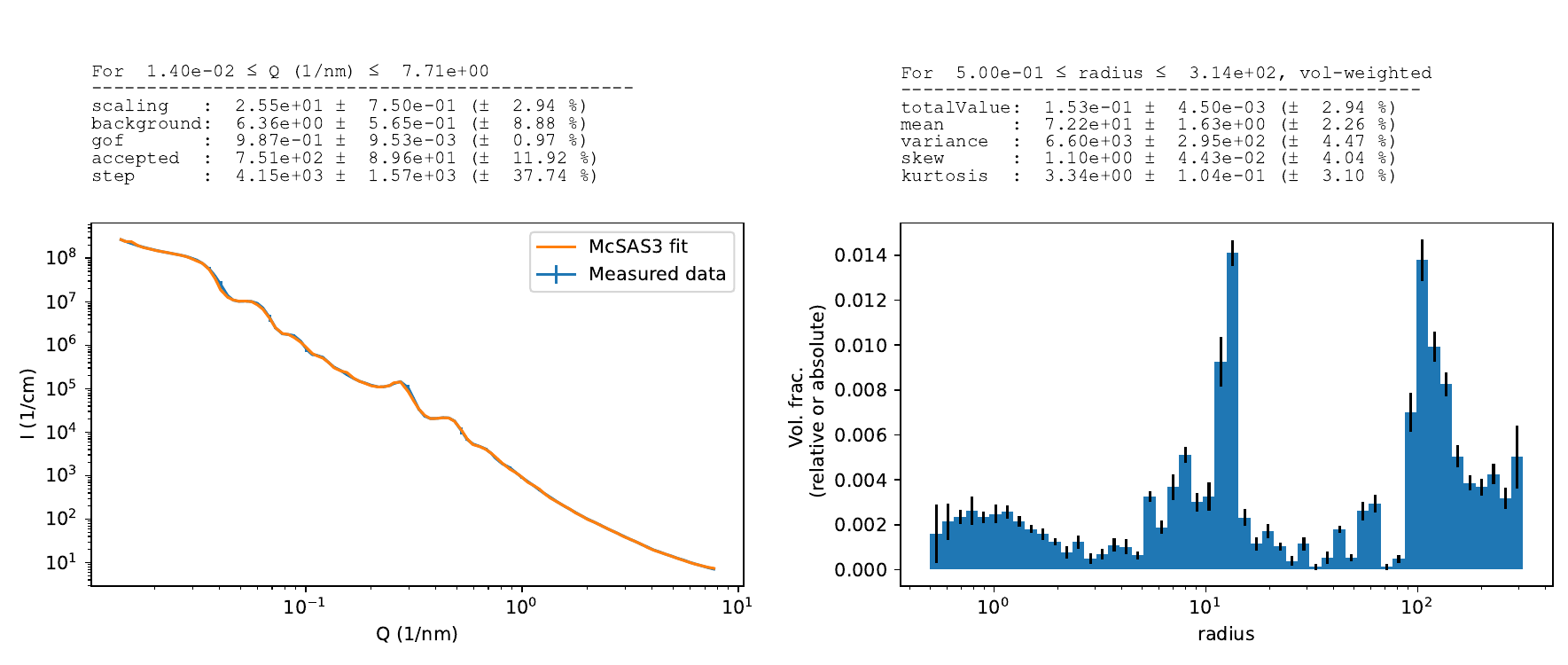}
\includegraphics[width=0.99\textwidth,trim={31cm 0 0 0},clip]{images/round_robin_dataset_3_output.pdf}
\caption{The analysis result of scattering from a dense, bimodal silica powder. Top left: data and fit. Top right: overall size histogram. Lower left: first population. Lower right: second population. }
\label{fg:ex2}
\end{center}
\end{figure}

This sample is a bit more challenging to analyze than Example 1, but with the right settings, we arrive at valid solutions. Note how the size distribution is affected by the configured volume fraction (valid answers can be obtained with a structure factor volume fraction parameter ranging from $\approx 0.46$ to $0.6$). This highlights that it is not possible to get both volume fraction and size distribution uniquely, but must fix one to retrieve the other as they are inextricably linked. To obtain the result in Figure \ref{fg:ex2}, we chose 0.55 because of speed: it is one of the fastest settings. The overall resulting distribution as well as the two isolated populations can be seen in Figure \ref{fg:ex2}

\subsection{Use example 3: Using a custom simulated form factor on a dispersion of faceted cubes}\label{ssc:ex3}

The third example demonstrates how McSAS3 can be used for shapes without analytical scattering functions. For these cases, simulated scattering patterns can be used as a basis model. 

The European Metrology Programme for Innovation and Research (EMPIR) ``nPSize'' project set out to make non-spherical reference materials. These narrow-dispersity, faceted cubes are sample number 4 from that project. As no analytical models for faceted cubes exist, the analysis with McSAS3 can be done using a model interpolated from a single scattering simulation of a faceted cube \cite{Deumer-2022a}. The resulting parameter distribution will then be the (uniform) scaling factors for this model object. 

Measurements of these nanoparticles were performed at the SAXS beamline at the PTB. The data is publicly available from Zenodo \cite{Deumer-2022}. The form factor model is a special ``sim''-type, which requires the simulated arrays for $Q$, $I$ and $\sigma_I$ of the form factor data to be specified in the model description. This model should be complete, and include a (flat) Guinier region at the start, and Porod region at the end. Parameters for extrapolation beyond the data should be explicitly specified (universally applicable, automated extrapolation parameter determination has not yet been implemented at the time of writing). 

The faceted cube data was simulated using the SPONGE, a software tool for simulating scattering from complex structures using the Debye equation, also described in \citeasnoun{Deumer-2022a}. The model was an STL file of a faceted cube with a face-to-face distance of 1 nm. This means the scaling factors in the final histogram are identical to the object size distribution in nm. 

The result from the McSAS3 optimization using this form factor is the set of scaling factors of the objects that, as an ensemble, best fit the data. This ``sim'' model fit assumes that a uniform scaling of the object results in a volume-squared intensity scaling and an inverse Q-axis scaling. The ``sim'' model can work in absolute units if the form factor model data itself is supplied in absolute units. 

\begin{figure}
\caption{The analysis result of scattering from suspension of faceted cubes. Left: data and fit, right: isolated population showing the size distribution of faceted cubes.}  
\includegraphics[width=0.99\textwidth]{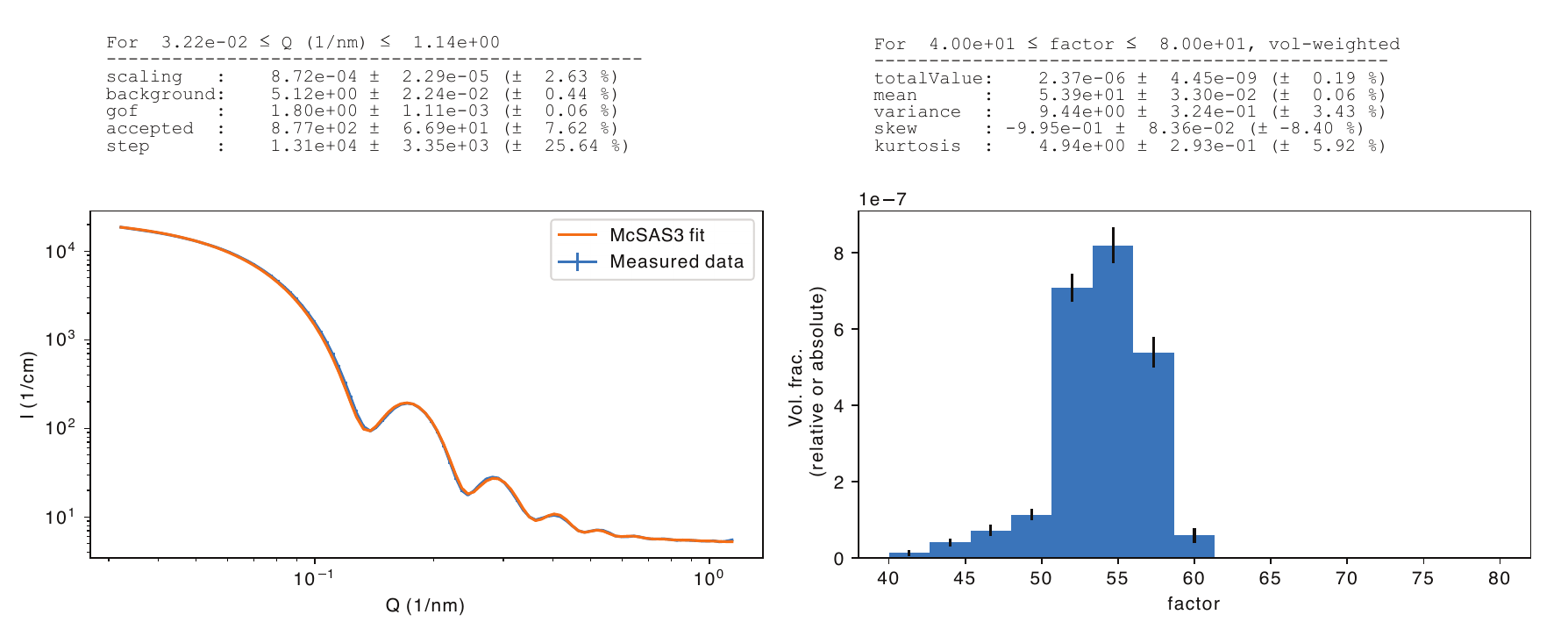}
\label{fg:ex3}
\end{figure}

Due to the very small data uncertainties, scaling factor range and/or discrepancies between model and reality, a full convergence to a reduced chi-squared of 1 cannot be obtained but can be approached. The result is shown in Figure \ref{fg:ex3}. 

\section{Adapting McSAS3 for lab and synchrotron data pipelines}

Given the increased quantities of data from both synchrotron beamlines and laboratory sources, the burden of data analysis on the scientists becomes even more pressing. Simultaneously, we are seeing a trend towards more in-situ and operando experiments even in the lab. Some of us also aim to integrate X-ray scattering as a part of autonomous material optimization feedback cycles \protect\footnote{Despite the nomenclature, in real life these automated set-ups actually require a large investment in highly-skilled personnel for the continuous integration, programming and operation.}. These trends strongly benefit from integrated, flexible data pipelines with analyses, if for nothing else than to gain an initial insight in the trends and developments. McSAS3 has been written largely for this purpose. 

To encourage users to use the software, the instrument responsible can make the data read-in configurations available for their instrument or data source. Likewise, a small library of optimization templates can be prepared which optimize with simple, near-universal models (spheres being the most practical example, either with structure factor for powders, or without for most other samples). These should be configured so that the optimizer completion criteria are mostly limited by 1) the maximum number of accepted moves, and 2) the maximum number of attempts, in order to keep the running time reasonable. It bears reminding here that the success of the optimizer is strongly dependent on reasonable uncertainty estimates on the data, and so a thorough data correction pipeline is essential \cite{Pauw-2017}. 

Once this is in place, this quick run of McSAS3 can be attached to the data processing pipeline, to provide users with (near-)immediate insight into their material development and to provide morphological data for feedback cycles and material optimizers to work with. Based on the results from the McSAS3 optimizations, a more artisanal analysis can be developed if required.

\section{Current limitations}

Software is never feature-complete, and McSAS3 is no different. There are a few limitations to bear in mind for now: 
\begin{itemize}
    \item The internal data model performs well but has no units support yet. Given the trouble we've seen where software users do not always pay attention to the ubiquitous units conversions our field unnecessarily demands \cite{Pauw-2023}, we need to make this more easy to the user. As we want to standardise on a new units- and uncertainties-aware data carrier developed for a modular data correction library under development, this improvement is already planned, but will require an in-depth change to the software foundation. As it is, the software is already very usable without this change. 
    \item Anisotropic data is supported in principle. The underlying fitting engine can take 1D or 2D data and fit either. The limiting factor for 2D data is the lack of support in the visualisation and graphical user interface. This can be implemented if a common use case surfaces that is worth the effort. 
    \item Lastly, there is no mechanism in place to interrupt an optimization, and if settings are poorly chosen, it can run for a very long time. The implementation of such functionality will require some thought and testing as the parallelisation and UI blocking makes interactive signaling more involved. For now, the recommendation is to start doing (test) optimizations only after carefully selecting appropriate settings for the maximum number of iterations. 
\end{itemize}

\section{Conclusions}

The hard limitations imposed by the original implementation of McSAS have now been resolved through a complete rewrite in ``McSAS3''. McSAS3 is intended to be integrated as part of an automated data processing pipeline, and the additional user interface ``McSAS3GUI'' provides an easy interface both for configuring the McSAS3 core, as well as for fitting smaller batches of data (e.g. several thousands of datafiles). The examples demonstrate the broad utility of the approach and implementation, as shown for bimodal particle dispersions, bimodal powders, and dispersions of scatterers with uncommon shapes. While the software will steadily improve in the future, the current state will undoubtedly be useful to many. 

\ack{Acknowledgements}

\referencelist{iucr}





\end{document}